\documentclass[aps,prl,reprint]{revtex4-1}
\usepackage{blindtext}
\usepackage{graphicx}
\usepackage{graphics}
\usepackage{bm}        % for math
\usepackage{verbatim}   % for math
\usepackage{amsfonts}
\usepackage{amsmath}
\usepackage{bm}
\usepackage{amssymb}
\usepackage{subfigure}
\usepackage{adjustbox}
\usepackage{lipsum}

\usepackage{adjustbox}

\usepackage{stfloats}

\usepackage{amsmath,amscd}
\usepackage {extarrows}
\usepackage{algorithm,algpseudocode}
	
\makeindex

\begin{document}

\title {Dynamics of Interacting Hotspots - I}

\author{Suman Dutta $^{\ast}$ }

\affiliation{The Institute of Mathematical Sciences, \\  IV Cross Street, CIT Campus, Taramani, Chennai,\\ Tamilnadu 600113.\\}

\email{Correspondence to: sumand@imsc.res.in}

%\footnote {$^{\ddag}$ Contributed equally}

\date{~\today}

\begin{abstract}
The worldwide spread of COVID-19 has called for fast advancement of new modelling strategies to estimate its unprecedented spread. Here, we introduce a model based on the fundamental SIR equations with a stochastic disorder by a random exchange of infected populations between cities to study dynamics in an interacting network of epicentres in a model state. Although each stochastic exchange conserves populations pair-wise, the disorder drives the global system towards newer routes to dynamic equilibrium. Upon controlling the range of the exchange fraction, we show that it is possible to control the heterogeneity in the spread and the co-operativity among the interacting hotspots. Data of collective temporal evolution of the infected populations in federal states of Germany validate the qualitative features of the model.

\end{abstract}

\maketitle

%%%MAIN TEXT%%%%

The sudden outbreak of the pandemic of the Novel Coronavirus (COVID-19) has caught emergency attention for its unprecedented rage in most parts of the globe, overwhelming the healthcare facilities, locking down cities, hitting hard the economy \cite{covid1, lancet1}. Within the first four months from the very first report from the city of Wuhan at the end of December 2019\cite{covid1,covid2}, the epidemic has already reached more than 200 countries, in all the inhabited continents, infecting more nearly 2.5 million people globally \cite{who1}. The death toll is already touching the sky at a devastating pace\cite{who1}.

Such pandemics, however, are not new in recent times: The 1918 Spanish flu resulted in 50-100 million deaths throughout the globe\cite{spanishflu,spanishflu2}. In recent years, similar coronavirus outbreak happened in 2003 (SERS-Cov) \cite{sars} and 2012 (MERS) \cite{mers} killing around a thousand people in China and the Middle east. Following that the outbreak of Ebola cost more than ten thousand lives in Africa\cite{ebola}. After resulting in massacres in Europe, the pandemic has turned its head towards the United States where more than half of a million people have already been infected within a short span\cite{worldometer}. The complexity in understanding and estimating the spatial spread of the pandemic lies in the conflicting numbers of the rates at which the virus is transmitted to the neighbourhood\cite{lancet2}. Specifically, when multiple hotspots coexist together, only a few recursive events of local transmission turn into a catastrophic avalanche with dramatically increasing number of infections in a very short time window\cite{prl1}. Moreover, the adaptability of the virus in ever-changing environments is also surprising\cite{virulant1}. Thus, identification of the true character and the self-organization of the epidemic has emerged as one of the biggest global challenges in human history. This is pivotal to develop non-pharmaceutical antidotes with sophisticated intervention strategies to control the dispersal of the pandemic until some clinical solutions arrive in support\cite{imp1}. 

Indeed, there are many major challenges in predicting the spatial spread in a country having multiple epicentres of human transmission, using phenomenological models\cite{rmp}. Since an effective model requires the prior knowledge of transmission and recovery rates, it is almost impossible to estimate them in the midway of an epidemic, for the rates depend on the nature of instantaneous social preparedness that vary locally. For instance, Germany had three major infected regions, Bavaria, Baden-Wuerttemberg and North Rhine-Westphalia which had roughly twenty thousand cases in each of these federal states on average by the first week of April\cite{rki}. In contrast, the other neighbouring states, Rhineland-Palatinate, Saarland, Hesse had surprisingly smaller numbers\cite{rki}. This situation is significantly different from Hubei that alone had more than eighty percent of the total cases in China\cite{who2}.

"Essentially, all models are wrong, but some are useful"\cite{box}. Here, we address the dynamics in an epidemic using a model of interacting hotspots. We consider a model state consisting of multiple cities that exchange infected populations pair-wise stochastically where the local dynamics in the individual cities are governed by the deterministic Susceptible-Infected-Recovered (SIR) model with instantaneously or regularly updated populations. Moreover, the order of the random exchange of the infected populations is controlled by a connectivity parameter. We show that tuning the connectivity parameter, it is possible to control the cooperativity of the infection propagation. However, the heterogeneity in the cooperativity determines controls over the outbreak. We validate the qualitative features of the model with the data of the heterogeneous evolution of COVID-19 in different provinces of Germany. We discuss further possibilities of development.

We consider model dynamics in a state consisting of different cities that are allowed to exchange populations among themselves. The population exchange is subject to a containment measure modelled via a connectivity parameter which is assumed to be a constant and identical for all cities. $M$ such interacting cities are considered with initial total populations $N^{i}$ being the sum of the Susceptible ($S^{i}$), Infected ($I^{i}$) and Recovered ($R^{i}$) populations in a city $i$. For simplicity, the exchange process uses the criteria of classic random exchange of econophysics\cite{econophysics}. During an exchange each of two cities retains an identical fraction of population $\eta$ and redistributes a random fraction of the infected population pair-wise. The interactions between a pair of cities ($i$ and $j$) with infected population $I^{i}$ and $I^{j}$ are governed by the exchange rules :

\begin{equation}
    I^{i}\rightarrow \tilde{I^{i}}
\end{equation}

\begin{equation}
    I^{j}\rightarrow \tilde{I^{j}}
\end{equation}

where

\begin{equation}
    \tilde{I^{i}}=\eta I^{i}+f_{R}(1-\eta)(I^{i}+I^{j})
\end{equation}

\begin{equation}
    \tilde{I^{j}}=\eta I^{j}+(1-f_{R})(1-\eta)(I^{i}+I^{j})
\end{equation}

with $I^{i}+I^{j}=\tilde{I^{i}}+\tilde{I^{j}}.$ The process also ensures the conservation: $N^{i}+N^{j}=\tilde{N^{i}}+\tilde{N^{j}}$ during the exchange processes. However. after $t=0$, the model does not use any restrictions to enforce any conservation in the global population. Here, $f_{R}$ is the random fraction (taken as a uniform deviate between 0 and 1) and $\eta$ is the connectivity parameter. The parameter serves in the same way as the saving propensity does in the context of econophysics models\cite{econophysics}. 

Now in the city, $i$, the time dependent population, $N^{i}$ is sum of instantaneous updated susceptible($S^{i}$), infected($\tilde{I^{i}}$) and recovered ($R^{i}$) populations following the $SIR$ equations of motions\cite{rmp}: 

\begin{equation}
  \frac{dS^{i}}{dt}=-\frac{\beta^{i}}{N^{i}}S^{i}\tilde{I^{i}} \label{eq1}
\end{equation}

\begin{equation}
    \frac{d \tilde{I^{i}}}{dt}=\frac{{\beta^{i}}}{N^{i}}S^{i}\tilde{I^{i}}-\gamma \tilde{I^{i}} \label{eq2}
\end{equation}

\begin{equation}
    \frac{dR^{i}}{dt}=\gamma^{i} \tilde{I^{i}}\label{eq3}
\end{equation}

We consider $\frac{M}{2}$ number of stochastic random exchanges using Eq. (3-4) between the cities before we update populations using Eqs. (5-7) (See numerical scheme in the Supplementary Information(SI), SI Note.1 ) in each time step. The number of exchanges are optimised in order to ensure significant amount of exchange of infected population among the cities that drives the process. In each of the time steps, we now solve $3M$ number of coupled differential equations alongside $\frac{M}{2}$ \cite{footnote0} number of random exchanges to extract the dynamics. We use the time step $\delta t=1$ for the integration throughout the calculation.

We monitor the dynamics of such a state consisting of $M=64, 1024$ and $4096$ cities with identical initial population ($N^{i}=1$ at $t=0$) for different ranges of magnitudes of connectivity parameter, $\eta$. Here $N^{i},S^{i},I^{i},R^{i}$
represent the fraction of the total population in the state $M$. Each city has identical initial infected population as $I^{i}(0)=0.0001$\cite{footnote1}. The initial conditions make sure that every city has identical behaviour for $\eta=1$ when the cities are made isolated by some intervention protocol as, for $\eta=1$, the cities retain all the population and does not actively take part in the exchanges. In contrast, they exchange freely for $\eta=0$.

\begin{figure}[h]
\includegraphics[scale=0.5]{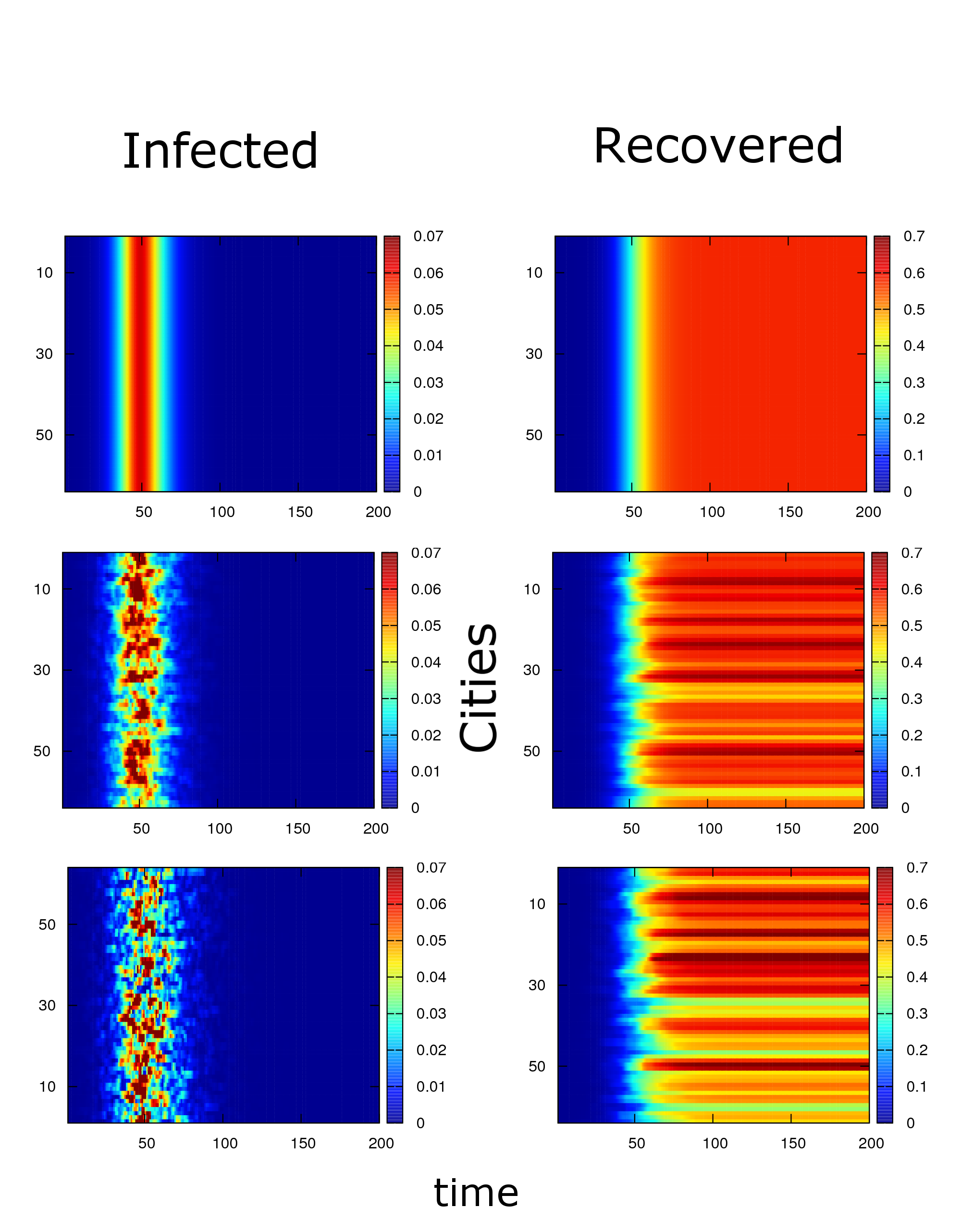}
\caption{Evolution of fraction of Infected (I) (Left Panels) and Recovered (R) (Right Panels) populations in a model state consisting of 64 cities with identical initial population, transmission rate($\beta=0.5$) and recovery rate ($\gamma=0.33$) for different connectivity indices ($\eta$), $\eta=1.0$ (Top Panels), $0.5$ (Middle Panels) and $0.0$ (Down panels). Dummy city labels are shown in y axis.}
\label{fig_entropy_disorder}
\end{figure}

In Fig. 1, we show the dynamics of the infected (I) and recovered (R) population in the state with $M=64$ for $\eta=1.0$, 0.5 and 0.0. For $\eta=1.0$, the cities are isolated completely. Hence, the behavior of the individual cities are identical to the solutions of SIR equations for a case with identical parameters (See SI Note:2 and SI Fig. 1). For $\beta =0.5$ and $\gamma =0.3333$, the peak in $I$ appears at $t\approx 50$ with a peak height $I^{P}=0.06$ while $S$ and $R$ saturates to steady values $S^{S} \approx 0.45$ and $R^{S}\approx 0.55$ respectively beyond $t\approx 80$. This we also see in the upper panels in Fig. 1 for $\eta=1.0$. The peaks in $I$ also appear simultaneously for all the cities $t\approx 50$.  

\begin{figure}[h]
\includegraphics[scale=0.09]{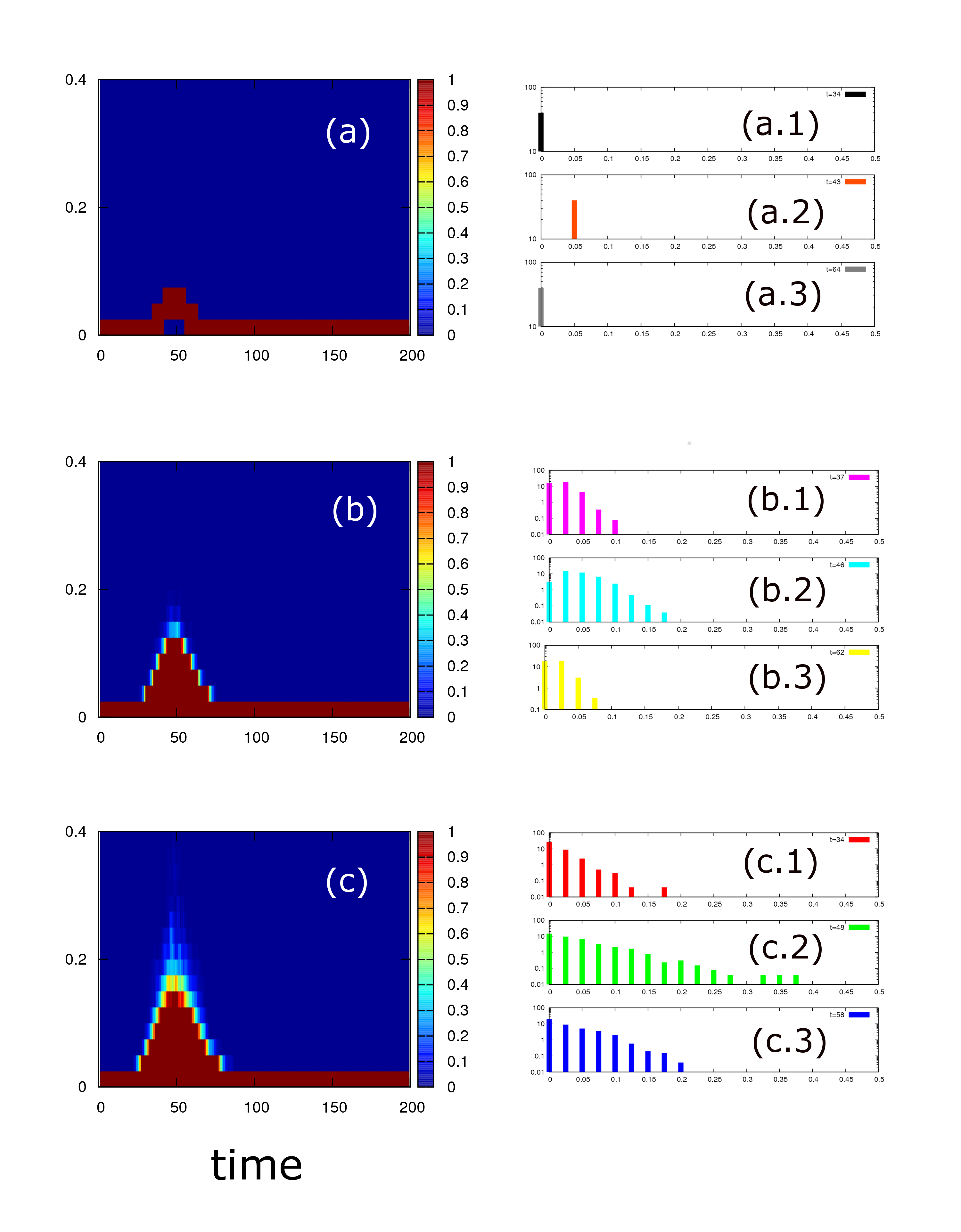}
\caption{(a-c) Time evolution of the probability distribution of infected populations in a city, $P(I^{i},t)$ vs infected populations in a city, $I^{i}$ with $t$ for different connectivity indices (a)$\eta=1.0$ (b)$\eta=0.5$ (c)$\eta=0.0$ in a state of $M=1024$ cities. Instantaneous $P(I^{i})$ vs $I^{i}$ at few $t$ is shown in subplots (a.1-3,b.1-3 and c.1-3).}
\label{fig_entropy_disorder}
\end{figure}

For $\eta < 1$, as the exchanges of the infected fraction between the cities set in, the behavioural differences in the cities become prominent. In the middle panels of Fig. 1, we show the evolution of Infected and Recovered population for $\eta=0.5$. We see heterogeneity in the infection propagation: Not only the peaks appear in different times in the different cities, but also the height of the peaks in $I^{i}$ become distinct for each city. The dynamic equilibrium is attained when $I^{i}\rightarrow 0$ for all $i$ simultaneously at sufficiently long time after the individual peaks in the infected populations in the cities are observed. These effects become more prominent for lower $\eta$. For $\eta=0$, the exchange is completely uncontrolled as individual cities do not have any restriction in retaining the fraction of infected population. Hence, the distributions of both $I$ and $R$ become more diffused as shown in the bottom panels of Fig. 1.

Now for all these cases, we look at the time evolution of the probability distribution of the infected populations, $P(I;t)$ with increasing $t$. In Fig. 2, we show the evolution of $P(I;t)$ for different cases of $\eta$. As there is a tiny population of infected population in the beginning due to the initial condition, the distribution at $t=0$, $P(I;0) \approx \delta (I(0))$ for all values of $\eta$. For $\eta=1$, as individual cities show a peak at $t\approx 50$, the delta peak of $P(I)$ also shifts to higher values of $I$ (Fig. 2.(a.2)). Since all the cities behave identically for $\eta=1$, there is no spread in the distributions for all $t$. At higher $t$ when the infected population $\rightarrow 0$, the distribution again shifts to smaller values of $I$(Fig.2(a.3)).

\begin{figure}[h]
\includegraphics[scale=0.09]{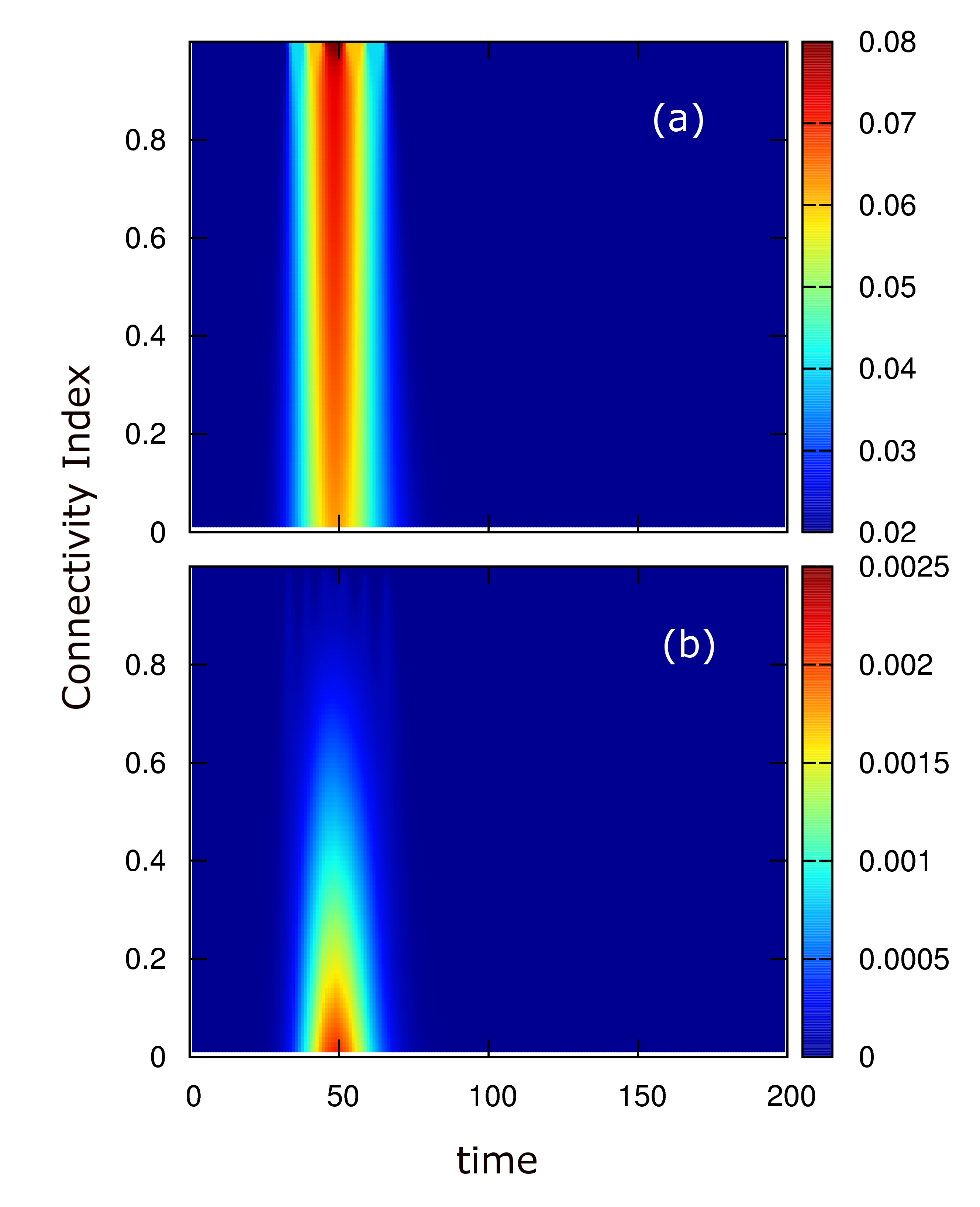}
\caption{(a) Phase diagram of the (a) mean and (b) Variance of the infected populations in a state consisting of $M=4096$ cities with varying connectivity index ($\eta$) and time ($t$) for identical transmission rate$ (\beta=0.5$), and recovery rate, $\gamma=0.33$. }
\label{fig_entropy_disorder}
\end{figure}

The situation changes for any $\eta < 1$ as the exchanges bring broadening in $P(I;t)$, specifically at $t\rightarrow 50$. For $\eta=0.5$, there is a significant broadening in $P(I)$ for $40 < t< 60$ [Fig.2(b)]. The nature of the tail in $P(I)$ albeit changes with $t$. We even observe a shift in the mode of the peak in $P^{i}$ [Figs. 2(b.1) and 2(b.2)]. The length of the tail decreases [Fig. 2(b.3)] and disappears at higher $t$. For $\eta=0$ [Fig.2(c)], the broadening in the distributions is more prominent, even at intermediate $t$[Fig.2(c.1)]. At $t\approx 50$, the tail in $P(I)$ exists upto $I\approx 0.375$. It affirms that there is finite probability to have cities with extremely large proportion of the infected populations of the order of $I\approx 0.375$ which is almost 37.5 percent of the total population [Fig.2(c.2)]. With increasing $t$, the tail decreases in $I$ and the peak reaches $\approx 0$ as the epidemic mitigates around $t\approx 75$. 

In order to understand the role of heterogeneity in the infection propagation, we now focus on the behaviour of the moments of $P(I)$ for different values of the connectivity index, $\eta$, at different $t$ [Fig. 3]. The mean of $P(I)$, $\mu= \int IP(I)dI$ is shown in Fig. 3(a). The behaviour of $\mu$ doesn't show much heterogeneity, suggesting that the collective peak appears almost at similar times as for individual cities. This doesn't change much as $\eta$ is varied. However, the height of the collective peak decrease for smaller $\eta$ around $t\approx 50$. The variance of $P(I)$, $\sigma^{2}(t)=<I(t)>^{2}-<I^{2}(t)>$ show interesting pattern for $t\approx 50$ [Fig. 3(b)] as $\eta$ is varied. Quite naturally, $\sigma^{2}(t)\rightarrow 0$ for $\eta\rightarrow 1$. As $\eta$ is decreased, the cooperativity also decreases due to increasingly larger disorder. For small $\eta$ and $\eta=0$, $\sigma^{2}(t)$ show prominent non-monotonic behaviour with increasing $t$. However, the non-monotonic dependence vanishes as $\eta\rightarrow 1$.

\begin{figure}[h]
\includegraphics[scale=0.32]{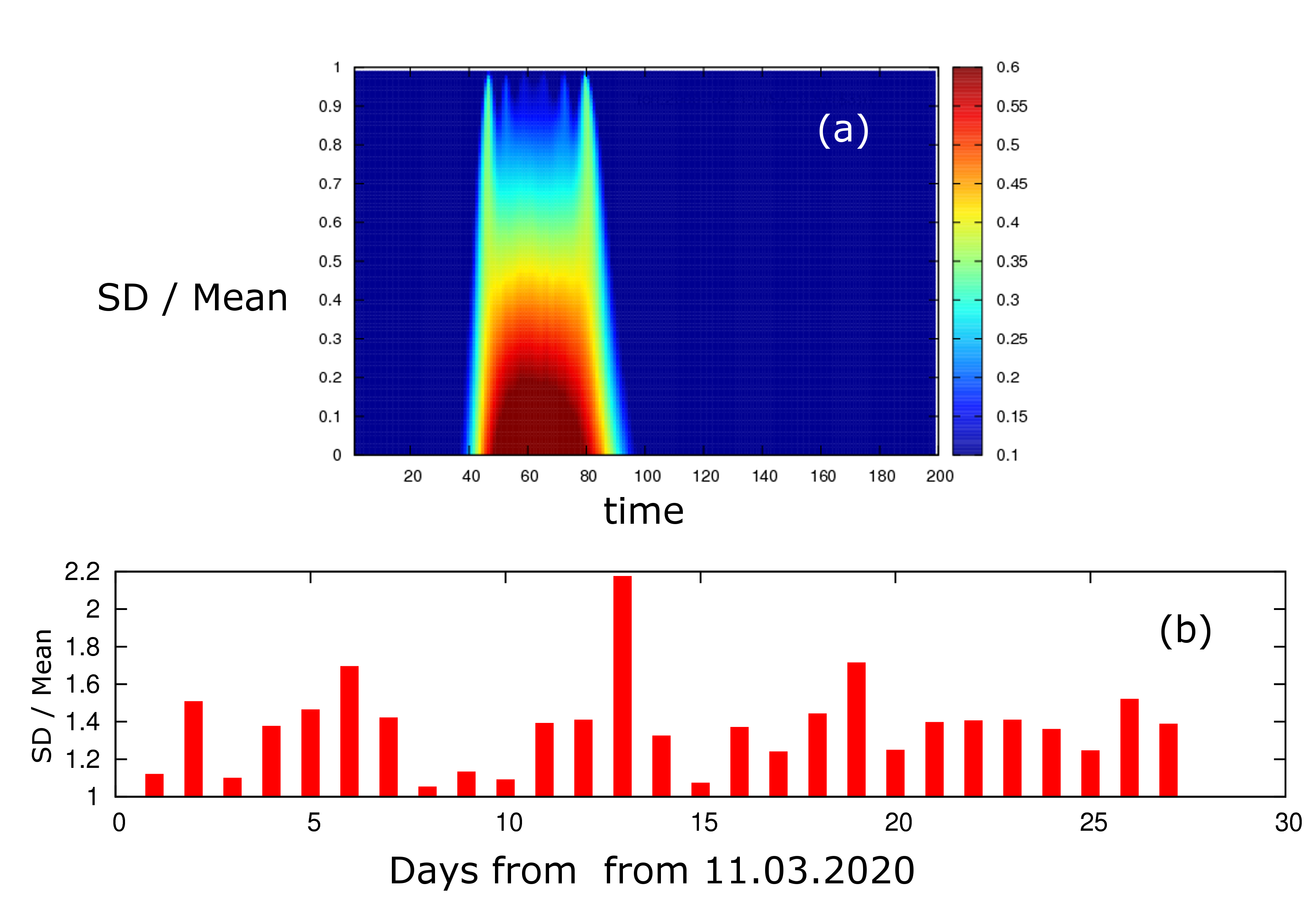}
\caption{(a) SD/Mean, $\sigma/\mu$ of the infected populations in a state consisting of $M=4096$ cities with varying connectivity index ($\eta$) and time ($t$) for identical transmission rate$ (\beta=0.5$), and recovery rate, $\gamma=0.33$. (b)$\sigma/\mu$ from the distribution of infected populations of COVID-19 from different federal states of Germany.}
\label{fig_entropy_disorder}
\end{figure}

The shift in $\mu$ occurs due to the simultaneous increase in the infected populations while the peak in $\sigma^{2}$ is due to distinct behaviours of the cities. We now look at the behaviour of $\sigma/\mu$ that captures the competition between the two factors. In Fig. 4(a), we show the phase diagram of $\sigma/\mu$. $\sigma/\mu$ has prominent non-monotonic dependence near $t\approx 50$ for smaller values of $\eta$. The peak defines the timescale of heterogeneity in the collective infection transmission.

\begin{figure}[h]
\includegraphics[scale=0.36]{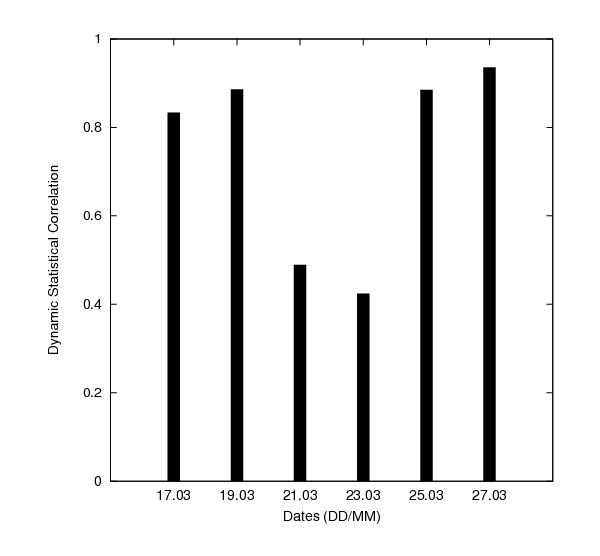}
\caption{Dynamic Statistical Correlation, $\rho (t,t+1)$ between the infected populations of COVID-19 from all the federal states in Germany at different dates in the month of March, 2020. }
\label{fig_entropy_disorder}
\end{figure}

We now attempt to understand the COVID-19 data of the cases in different provinces of Germany using the insights of the model. The time series data for Germany for all 16 provinces are shown in Supplementary Information (SI Figs. 2-5). We now analyse the behaviour of $\sigma/\mu$ computed from the distribution of the infected populations in provinces of Germany (See SI Figs. 2-5). We see $\sigma/\mu$ has a sharp peak on 23rd March, 2020. In order to verify the heterogeneity, we further look at the dynamic statistical correlation between the daily new cases in the different provinces for the successive days, using the Pearson correlation between the distribution of the infected populations at all the provinces in time $t$ and $t+1$, given by

\begin{equation}
    \rho (t,t+1)=\frac{cov[P(I;t), P(I;t+1)]}{\sigma(t)\sigma(t+1)}
\end{equation}

We show the dynamic statistical correlation, $\rho(t,t+1)$ between the number of cases on successive days from all the federal states of Germany in Fig. 5. We avoid the double count in the data. The trend in the statistical correlation behaves non-monotonically in the period, with a dip on 23rd March when the peak in $\frac{\sigma}{\mu}$ appeared, reiterating the effect of heterogeneity due to lack of dynamical statistical correlation. Surprisingly, the date exactly matches the date of the beginning of the lockdown in Germany. From the first week of April, there is a clear trend of {\it flattening of the curve} with significantly smaller number of cases with respect to the maximum numbers recorded on 27th March.

Always, there are continuous exchanges of populations between the provinces in a country unless any intervention is imposed. The mobility between the cities without restriction, in principle resembles the $\eta=0$ case of the model while the situation with strict intervention behaves as $\eta=1$ case of the model when the cities are considered isolated. Considering time dependence in $\eta$, it is possible to realize a dynamic picture from pre to post lockdown situations. The behaviour of $\sigma/\mu$ qualitatively features the balance between the spread of the infection in the cities and the relative change in the collective spread of the epidemic in the state. The extremum in $\sigma/\mu$ captures the extreme heterogeneity in the infection propagation as it optimize the relative balance between the local spread and drive in the collective transmission. On this note, the model that we use delivers the similar qualitative feature of the data of COVID-19 cases in the states of Germany. 

Whether the appearance of the peak in heterogeneity is a universal confirmatory signature of containment, can be confirmed only by further studies. There are many possible ways to develop the minimalistic model. It is possible to include time dependence in the parameters. Random deviates of a different kind can also be tested for modelling specific situations. Also, possibilities of further customization needs system specific details and the set of governing equations where specific rules of death, quarantine or hospitalization counts could be considered in addition. Further, it would be challenging to model the transportation of the infected populations using deterministic equations\cite{ds}, vis-a-vis  the data driven approaches\cite{science3, science4}. On this note, we believe, that this study, opens up horizons in exploring the rich dynamics of infection propagation inluding sophisticated intervention among the interacting hotspots.

 The author is indebted to D. Syam for continuous encouragements and numerous insightful discussions, P. Das for technical support and P. Chaudhuri for comments on the draft. The data of cases of COVID-19 for the federal states of Germany was taken from the public archive of the situation reports of the Robert Koch Institute on COVID-19 at www.rki.de. There are no conflicts to declare.


\begin{thebibliography}{}
\bibitem{footnote0}
Considering a single population exchange in each timestep equilibrate things ideally. However, this is an extremely slow process and realistic scenario could be simulated using a smaller value of timestep $\sim \frac{\delta t}{M}$. We speed up using $\frac{M}{2}$ pair-wise exchanges in each timestep before we integrate the epidemic variables deterministically after every city takes part in the exchange process at least for once, probabilistically in every timestep.
\bibitem{footnote1}
In principle, there is no restriction on the initial conditions. We tested that the equations work for many set of initial conditions that includes cases with cities with arbitrary distribution of initial population in the cities and suitable choices of transmission rates and recovery rates.
\end{thebibliography}
\end{document}